\documentclass[12pt,preprint]{aastex}
 
\begin{document}

\title{Detection of OH absorption against \pulsar}
 
\author{Anthony H. Minter}

\affil{National Radio Astronomy Observatory, Green Bank, WV, 24944}

\email{tminter@nrao.edu}

\newcommand{\pulsar}{\object[PSR B1718-35]{PSR B1718-35}}
\newcommand{\parkes}{\object[PSR B1641-45]{PSR B1641-45}}
\newcommand{\arecibo}{\object[PSR B1849+00]{PSR B1849+00}}
\newcommand{\ngc}{\object[NGC 6334]{NGC 6334}}

\newcommand{\srca}{\object[PSR B1736-31]{PSR B1736-31}}
\newcommand{\srcb}{\object[PSR B1750-24]{PSR B1750-24}}
\newcommand{\srcc}{\object[PSR B1809-176]{PSR B1809-176}}
\newcommand{\srcd}{\object[PSR B1815-14]{PSR B1815-14}}
\newcommand{\srce}{\object[PSR B1817-13]{PSR B1817-13}}

\newcommand{\srcf}{\object[PSR B0458+46]{PSR B0458+46}}
\newcommand{\srcg}{\object[PSR B1703-40]{PSR B1703-40}}
\newcommand{\srch}{\object[PSR B1737-30]{PSR B1737-30}}
\newcommand{\srci}{\object[PSR B1818-04]{PSR B1818-04}}
\newcommand{\srcj}{\object[PSR B1829-08]{PSR B1829-08}}
\newcommand{\srck}{\object[PSR B1834-10]{PSR B1834-10}}
\newcommand{\srcl}{\object[PSR B1845-01]{PSR B1845-01}}
\newcommand{\srcm}{\object[PSR B2106+44]{PSR B2106+44}}
\newcommand{\srcn}{\object[PSR B2111+46]{PSR B2111+46}}
\newcommand{\srco}{\object[PSR B1648-42]{PSR B1648-42}}

\newcommand{\searcha}{\object[PSR B1737+13]{PSR B1737+13}}
\newcommand{\searchb}{\object[PSR B1933+16]{PSR B1933+16}}
\newcommand{\searchc}{\object[PSR B2016+28]{PSR B2016+28}}
\newcommand{\searchd}{\object[PSR B1944+17]{PSR B1944+17}}
\newcommand{\searche}{\object[PSR B1915+13]{PSR B1915+13}}
\newcommand{\searchf}{\object[PSR B1929+10]{PSR B1929+10}}
\newcommand{\searchg}{\object[PSR B0740-28]{PSR B0740-28}}
\newcommand{\searchh}{\object[PSR B0833-45]{PSR B0833-45}}
\newcommand{\searchi}{\object[PSR B0835-41]{PSR B0835-41}}
\newcommand{\searchj}{\object[PSR B0906-49]{PSR B0906-49}}
\newcommand{\searchk}{\object[PSR B1054-62]{PSR B1054-62}}
\newcommand{\searchl}{\object[PSR B1055-52]{PSR B1055-52}}
\newcommand{\searchm}{\object[PSR B1154-62]{PSR B1154-62}}
\newcommand{\searchn}{\object[PSR B1240-64]{PSR B1240-64}}
\newcommand{\searcho}{\object[PSR B1323-58]{PSR B1323-58}}
\newcommand{\searchp}{\object[PSR B1323-62]{PSR B1323-62}}
\newcommand{\searchq}{\object[PSR B1557-50]{PSR B1557-50}}
\newcommand{\searchr}{\object[PSR B1601-52]{PSR B1601-52}}
\newcommand{\searchs}{\object[PSR B1742-30]{PSR B1742-30}}
\newcommand{\searcht}{\object[PSR B1749-28]{PSR B1749-28}}
\newcommand{\searchu}{\object[PSR B1800-31]{PSR B1800-31}}
\newcommand{\searchv}{\object[PSR B1822-09]{PSR B1822-09}}
\newcommand{\searchw}{\object[PSR B1826-17]{PSR B1826-17}}
\newcommand{\searchx}{\object[PSR B0329+54]{PSR B0329+54}}
\newcommand{\searchy}{\object[PSR B0450+55]{PSR B0450+55}}
\newcommand{\searchz}{\object[PSR B0525+21]{PSR B0525+21}}
\newcommand{\searchaa}{\object[PSR B0740-28]{PSR B0740-28}}
\newcommand{\searchbb}{\object[PSR B1642-03]{PSR B1642-03}}
\newcommand{\searchcc}{\object[PSR B1749-28]{PSR B1749-28}}

\newcommand{\hi}{\ion{H}{1}}
\newcommand{\hii}{\ion{H}{2}}
\newcommand{\kms}{\,km\,s$^{-1}$}

\newcommand{\nraoblurb}{\footnote{The National Radio Astronomy Observatory is a
facility of the National Science Foundation operated under cooperative
agreement by Associated Universities, Inc.}}

\begin{abstract}

OH absorption against \pulsar\ at $(l,b) =351\fdg688, +0\fdg671$
has been discovered at 1665 and 1667\,MHz 
using the Green Bank Telescope.  The absorption appears to arise at
the interface of an \hii\ region and a molecular cloud
which are likely associated with the high mass star forming region \ngc.  
Beam dilution is
found to be the cause of differences in the opacity of the OH
against the Galactic background continuum emission and against the pulsar.  
The OH cloud is approximately 3 by 1.3\,pc and is located behind the 
\hii\ region.

\end{abstract}
 

\section{Introduction}

\hi\ absorption measurements against pulsars have provided a wealth
of information on the structure of the Interstellar Medium (ISM).  
They provide one of the best means of obtaining distance estimates to 
a large number of pulsars \citep[e.g.][]{frail}.  Combining the
distances with the
pulsar's dispersion measure (DM) allows the average electron density along
the line of sight to the pulsar to be determined.
This also allows models of the Galactic electron density
to be computed \citep[e.g.][]{cordes,lazio}.  The \hi\ 
absorption measurements essentially provide the means by which the electron
density model and the atomic neutral gas models of the Galaxy are tied 
together.  Comparison of the \hi\ emission when
the pulsar is ``off'' with the \hi\ absorption against the pulsar 
can provide information on the density and spin temperature of the 
absorbing \hi\ gas \citep[e.g.][]{balser}.
\hi\ absorption measurements against pulsars have 
been used to search for structure in the neutral 
atomic gas on very small scales \citep[e.g.][]{balser, tsas}.
These measurements also play an important role in the study of
pulsar population statistics, providing the spatial
distribution of pulsars throughout
the galaxy \citep[e.g.][]{population}.

It is only natural to attempt to extend absorption measurements against
pulsars to molecular gas.  Since pulsars have steep spectral indices,
it is necessary to look for molecules which have transitions at lower 
frequencies and which have relatively strong line strengths.  
This makes the ${\rm\bf ^2\pi_{3/2}(J=3/2)}$ transitions
of the hydroxyl radical (OH) at 1612, 1665, 1667 and 1720\,MHz
ideal for looking for absorption from molecular 
material against pulsars.
 
Previously, only two pulsars have been found
which have measurable OH absorption, \arecibo\ discovered by \citet{snez}
using the Arecibo radio telescope and \parkes\ discovered by \citet{joel}
using Parkes radio telescope.  Both \citet{snez} and \citet{joel} observed
numerous bright pulsars, 25 in total.  In both cases, the OH absorption
seen toward the pulsars have unique properties.  The OH
absorption against the pulsars have narrower linewidths and deeper absorption
(larger opacities) than is observed when the pulsar is ``off.''
This has been attributed to the presence of small scale structure in the
molecular material along the lines of sight to the pulsars.

There are a large number of pulsars which have not been observed for OH
absorption.  In this paper I present observations of sixteen pulsars 
(\srca, \srcb, \srcc, \srcd, \srce, \srcf, \srcg, \srch, \srci, \srcj, \srck,
\srcl, \srcm, \srcn, \srco, and \pulsar ) looking for OH
absorption.  OH absorption was only detected against \pulsar.

\section{Observations and Data Reduction}

The OH absorption measurements were made using the National Radio
Astronomy Observatory's (NRAO\nraoblurb ) 100~m Green Bank
Telescope (GBT).
The GBT has an unblocked aperture and a spatial resolution
of 7.4 arc-minutes at 1665~MHz.  The 1 to 2 GHz receiver, placed at
the focus of the Gregorian optics system,  was used
for the observations.  This receiver has a nominal bandpass of
$1100-1752$~MHz, dual linear polarization and had a system temperature on 
cold sky of
$18$ K.  The NRAO spectral processor, an FFT
spectrometer, was used to simultaneously observe the 1612, 1665, 1667 and
1720 MHz transitions of OH using orthogonal linear polarizations.  
The spectral processor
integration time was approximately 10 seconds, set to the nearest 
integer number of pulse periods.
The spectral processor's 32-level sampling provides excellent
dynamic range when Radio Frequency Interference (RFI) is present.
Most observations used 256 spectral
channels per linear polarization per OH transition with a bandwidth 
of 2.5~MHz, producing a spectral
resolution of $\sim 1.75$\kms\ per channel.  
Higher resolution observations were performed for \pulsar, using
256 spectral
channels per linear polarization per OH transition with a bandwidth 
of 0.625~MHz, producing a spectral
resolution of $\sim 0.44$\kms\ per channel.  

The dates and times of the observations are listed in Tables~\ref{table:obs}
and~\ref{table:obs1718}.  Table~\ref{table:obs} lists the observing dates
and times for those pulsars for which OH absorption was not detected.
Table~\ref{table:obs1718} lists the dates, times and bandwidths used for
the OH absorption observations toward \pulsar.
The pulsar OH absorption data were reduced using the technique outlined in 
\citet[][]{minter}.  The data were flagged for instances when RFI was present.

\pulsar's Galactic coordinates are $(l,b) =351\fdg688, +0\fdg671$.
Since \pulsar\ lies in the Galactic plane where continuum emission
is present, a slice at constant Galactic longitude was observed 
at the longitude
of \pulsar.  The observations went approximately $\pm 5^\circ$ out of
the galactic plane.  These observations allow the Galactic continuum
along with the continuum from a discrete object in front of \pulsar\
to be measured.  Combined with the pulsar ``off'' spectra, the 
opacities of the OH lines can then be calculated.
The data were corrected for the opacity difference between the zenith
and the low elevation (10$^\circ$ -- 12$^\circ$) of the observations.  
A zenith opacity
of $\tau=0.0103$ at 1666\,MHz was used along with an atmospheric temperature
of 250\,K.  The system temperature of $\sim15$\,K found near the zenith
was then subtracted from the data.    The remaining brightness temperature was
assumed to represent the continuum emission from the Galactic Synchrotron 
and free-free emission (assumed to be a smooth component)
and \hii\ regions or SNRs along the line of sight toward \pulsar.

\section{Source Selection}

The two prior large searches for OH absorption against pulsars by 
\citet{snez, joel} selected
their targets based on the brightness of the pulsars.
This yielded a detection rate of two out of twenty-five (8\%).  This motivated
me to look for a better selection criteria for pulsars with OH absorption.

\subsection{Scattering As A Possible Selection Criterion}

\citet{boldyrev} proposed that the strongest
Interstellar Scattering (ISS) of pulsar
signals was not induced by Kolmogorov-like turbulence in the electrons
within the ISM, but by refraction at the random interface between ionized 
and non-ionized gas along the line of sight.  Particularly, it was predicted 
that the Photo-Dissociation Regions (PDRs) at the edges of molecular clouds 
should produce the strongest scattering.  
\citet{boldyrev} also suggest that a pulsar's time broadening would scale as
density to the fourth power, $(\tau \propto n^4)$, so that only the
most dense PDR along the line of sight will dominate the ISS.
This will give the appearance of a single scattering screen toward
the pulsar which is consistent with what is observed \cite[e.g.][]{hill}.

I list prior attempts to observe OH absorption against pulsars
in Table~\ref{table:scattering}.  From Table~\ref{table:scattering} it is seen
that the only OH absorption found (including all possible detections)
are along lines of sight that are highly scattered and which have
large amounts of CO.  The pulsars with non-detections  
are typically bright pulsars, have relatively weak
scattering and little CO along their lines of sight.  
By just looking at bright pulsars, the searches for OH absorption have
been biased to lines of sight with little molecular material.  
If Boldyrev \& K\"{o}nigl's hypothesis is correct then the OH absorption
searches have also been biased since they tend to be toward weakly scattered
pulsars.

I thus created a list of the most highly scattered pulsars that had
a measurable amount of CO along their lines of sight.  I obtained the
scattering values from the ATNF pulsar catalog \citep{atnf} and the
CO information from \citet{co}.  I then selected
the pulsars that were outside the Declination range of Arecibo, viewable
with the GBT and which could reach a limiting OH opacity of $\tau \sim 0.1$
in less than ten hours as my source list.
This list included the pulsars \srca, \srcb, \srcc, \srcd, \srce\ and \pulsar\
which were observed in June -- October, 2005.

\subsection{OH Emission As A Selection Criterion}

Another approach that I took was to come up with a list of pulsars
that had detectable CO emission along their lines of sight.  I excluded
pulsars in the declination range of Arecibo, pulsars with Galactic 
latitudes greater than five degrees, pulsars that would take more than 
30 hours to reach $\tau\sim0.1$ and pulsars that had been previously observed
for OH absorption. There were 85 pulsars that met this criteria.
I then observed the OH emission toward these pulsars with the GBT
to determine which lines of sight had detectable OH 
\citep[these results will be presented in a future paper;][]{ohsearch}.

OH was detected on the lines of sight toward 34 of the 85 pulsars.
The velocity of the OH emission, along with the DM--derived distance
of each pulsar, was used to determine which pulsars could have OH
between the Sun and the pulsar.  This then led to observations of 
\srcf, \srcg,
\srch, \srci, \srcj, \srck, \srcl, \srcm, \srcn\ and \srco\ with the GBT
to search for OH absorption against these pulsars.

\section{Results}

No OH absorption was detected toward \srca, \srcb, \srcc, \srcd, \srce,
\srcf, \srcg, \srch, \srci, \srcj, \srck,
\srcl, \srcm, \srcn, and \srco.
The $1 \sigma$ upper limits for the opacity for OH absorption toward
these pulsars in shown in Table~\ref{table:limits}.  \srcb\ did not have
any detectable OH in the pulsar ``off'' spectra and was thus not observed
for a significant amount of time, resulting in its higher opacity limit.
OH absorption was only detected against \pulsar.

\subsection{PSR~B1718-35 OH Absorption Properties}

\begin{figure}
\epsscale{0.9}
\includegraphics[width=6.5in]{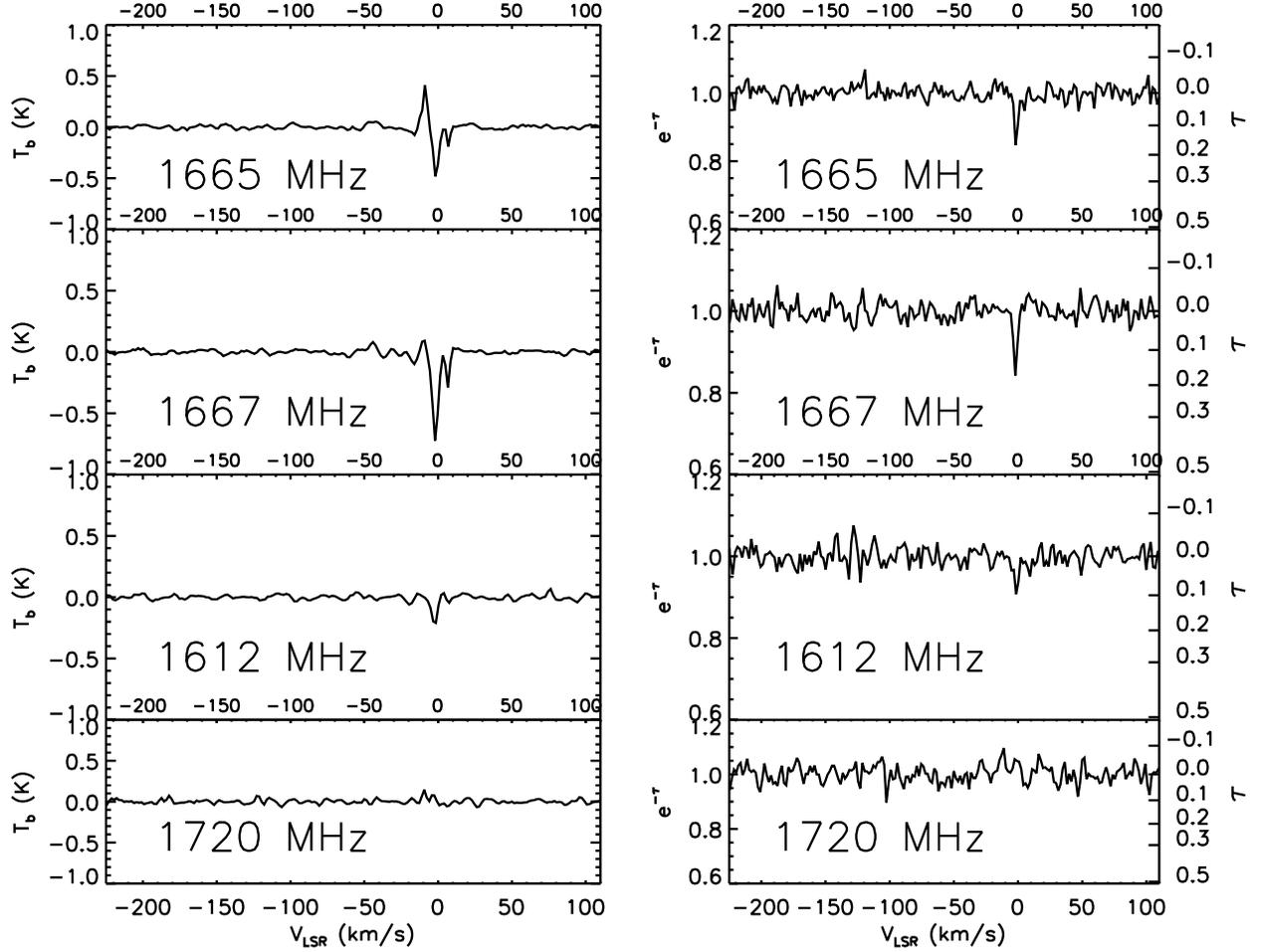}
\caption{The observed OH opacities observed against PSR~B1718-35 with 
1.75\kms\ spectral resolution.  The left column shows the
pulsar ``off'' spectra and the right column shows the OH absorption
against PSR~B1718-35.}
\label{fig:wide}
\end{figure}

\begin{figure}
\epsscale{0.9}
\includegraphics[width=6.5in]{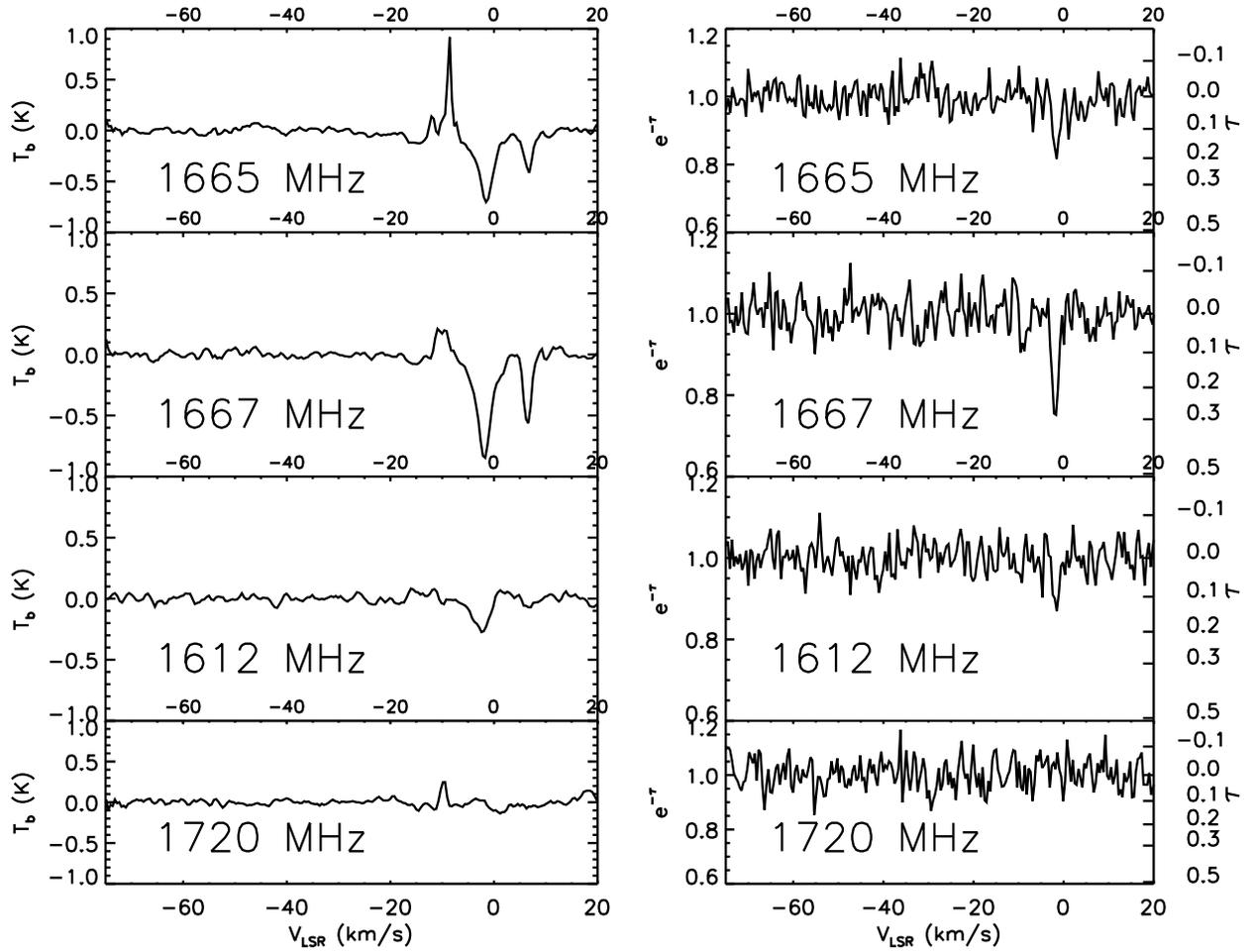}
\caption{The same as Figure~\ref{fig:wide} except for the higher 
0.44\kms\ spectral resolution data.}
\label{fig:narrow}
\end{figure}

I show the OH absorption spectra against \pulsar\ as well as the pulsar 
``off'' OH spectra for the two resolution modes (1.75 and 0.44\kms )
in Figures~\ref{fig:wide} 
and~\ref{fig:narrow}.  After initially detecting the OH absorption against 
\pulsar\ in the lower resolution mode, which allowed a large velocity range 
to be searched, I switched to the higher resolution mode to resolve the
narrow absorption lines that were present.  The results of Gaussian
fits to the OH absorption lines are presented in Table~\ref{table:datapulsar}.
As can be seen from the results in Table~\ref{table:datapulsar}, the initial
wide-bandwidth observations, which had a spectral resolution slightly larger
than the true linewidths of the spectral lines, are clearly affected by
poor spectral resolution.

\subsubsection{Column Density of the OH Absorbing Cloud\label{sec:column}}

From the observed opacity of the 1667\,MHz absorption against \pulsar, 
we can estimate the OH column density, $N_{OH}$, in the absorbing cloud.  
Using equation~9.12 of
\citet{elitzur}, along with $\tau(1667)=0.3\pm0.003$ and a linewidth
of $\Delta v = -1.5\pm0.2$, the OH column density is given by
\begin{equation}
N_{OH} =  {1.\pm 0.1 \times 10^{14} \over T_x(1667)} \,{\rm cm}^{-2}
\end{equation}
where $T_x(1667)$ is the excitation temperature of the line which
is typically 5-10\,K \citep{elitzur}.  The observed linewidth limits
the thermal temperature of the OH to be $T < 830\pm240$\,K, resulting
in the limit that $N_{OH} > 1.2\pm0.1 \times 10^{11}$\,cm$^{-2}$. 
From the typical excitation temperatures, the likely column density is
of the order $N_{OH} \sim 10^{13}$\,cm$^{-2}$.
For OH column densities between 
$10^{14} < {N_{OH} \over \Delta v}  < 10^{15}$\,cm$^{-2}$\,s\,km$^{-1}$
the satellite lines at 1612 and 1720 MHz are conjugates with the 1720\,MHz
line being in emission and the 1612\,MHz line being in absorption
\citep{joel}.  At column densities 
${N_{OH} \over \Delta v}  > 10^{15}$\,cm$^{-2}$\,s\,km$^{-1}$
the lines remain conjugates but with the 1612\,MHz line being in emission
and the 1720\,MHz line being in absorption.
Both of these cases occur when the region containing the OH is optically
thick to infra-red photons.  As can be seen in the pulsar ``off''
spectra in Figure~\ref{fig:narrow}, the 1612\,MHz line is in emission
at the velocity where absorption is seen against the pulsar.  However,
the 1720\,MHz line is not detected at a $3 \sigma$ limit of $0.08$\,K.
This is significantly less than the $-0.25$\,K that we would expect if
the 1720\,MHz line exhibited conjugate emission with the 1612\,MHz line.
This indicates that the OH column density is 
$N_{OH} < 6.7\pm0.9 \times 10^{13}$\,cm$^{-2}$ in 
the OH absorption absorbing cloud toward \pulsar.  Using the standard abundance
ratio of $N_{OH} / N_{H} = 6 \times 10^{-8}$ \citep{elitzur},the total hydrogen
column density in the absorbing cloud is  of the order
$N_{H} \sim (1.7-11.1) \times 10^{20}$\,cm$^{-2}$.  The \ion{H}{1} absorption
against \pulsar\ was measured by \citet{wsfj} and is shown in their
Figure~1d.  At the velocity of the OH absorption, the \ion{H}{1} absorption
has an opacity $\tau > 2$ and the emission has $T_B\sim 100$\,K.  Assuming
that half of the total \ion{H}{1} emission comes from the same side of the
tangent point\footnote{The velocity--distance relationship is doubled 
valued for the \pulsar\ line of sight.  Since the Galactic latitude
is small, $b=0\fdg671$, we can expect that the line of sight we encounter
equal amounts of gas on the near side and the far side of the tangent 
point.  Thus the total emission at any velocity will have half of its
contribution from the near side with the other half coming from the far
side of the tangent point.}
as the OH cloud results in an estimated column density 
$N_{HI} > 1.8 \times 10^{20}$\,cm$^{-2}$ which is consistent with the
column density derived from OH.

\subsubsection{The Distance To The OH Cloud}

\citet{wsfj} determined that the \ion{H}{1} kinematic distance to \pulsar\
is $5.6\pm 0.6$\,kpc.  This puts \pulsar\ on the near side of the
tangent point for its line of sight.  The Galactic bar is on the far
side of the tangent point for the \pulsar\ line of sight and can be
ignored in the following discussions.

\begin{figure}
\epsscale{0.9}
\includegraphics[scale=0.65, angle=90]{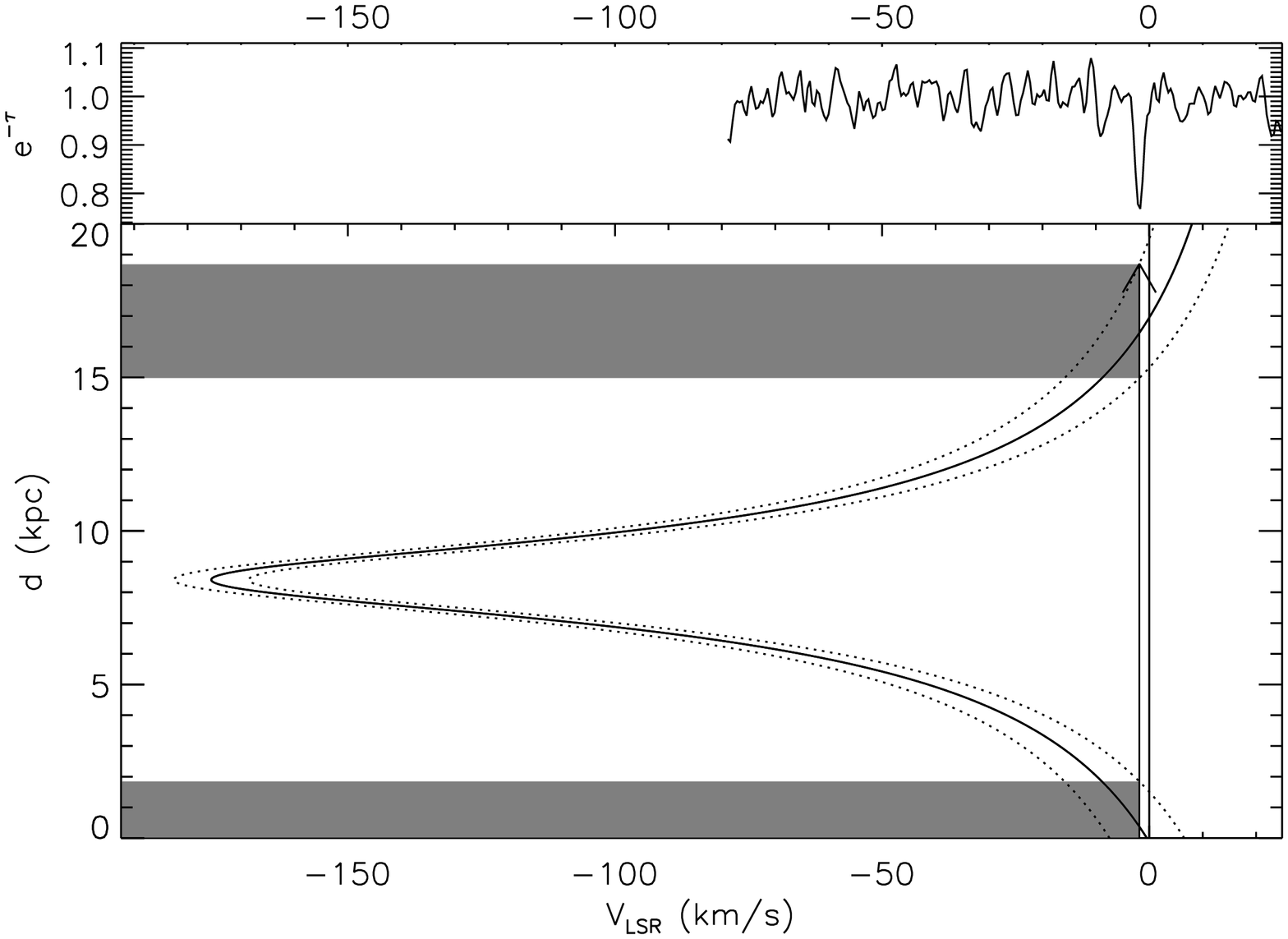}
\caption{The velocity--distance relationship for the flat rotation curve
of \citet{fich}.  The 1667\,MHz OH absorption line against PSR~B1718-35
is shown in the top panel.  The flat rotation curve is shown as the solid
line in the lower panel.  The dashed lines represent the random velocity
of 7\kms\ found for cold clouds \citep{jay}.  The vertical arrow
is at the center velocity of the OH absorption.  The gray shaded areas
show allowed distances for the OH absorption cloud.
The distance is limited to be within 1.84\,kpc of the Sun, given that 
PSR~B1718-35 is on the near side of the tangent point \citep{wsfj}.}
\label{fig:flat}
\end{figure}

I show the velocity--distance relationship for
the \pulsar\ line of sight using the flat rotation curve of \citet{fich}
In Figure~\ref{fig:flat}.
In determining the distance of the OH cloud from its velocity, a random
motion for cold clouds of 7\kms\ has been used \citep{jay}.
This random velocity was used as an error on the distance model as
suggested in \citet{minter}.  An upper limit of 1.86\,kpc is found for
the distance to the OH cloud.  

The line of sight toward \pulsar\ is in the extended halo of radio continuum 
emission surrounding \ngc\ \citep[see the Figure on page 347 of][]{altenhoff}.
Comparing the observed velocity range, $-1.5$ to $-1.9$\kms, for the OH 
absorption against \pulsar\ with the CO velocities of \cite{ngcco}
shows that the OH cloud has the same velocity as \ngc\,B and the extended
CO emission to the north-east of \ngc\ 
\citep[see Figure~6 of][]{ngcco}.  \hii\ regions
in and around \ngc\ have velocities in the range $+1$ to $-7$\kms\
\citep{jayhii, qrbb}.  It is thus quite possible that the OH cloud is part of
the \ngc\ Giant Molecular Cloud.  If this is the case then we can place
the OH cloud at the distance of \ngc, $1.7\pm0.3$\,kpc 
\citep{ngcdist}, which is consistent with the kinematic distance upper limit.

\subsection{The Continuum Emission Toward PSR~B1718-35 }

The observed continuum emission in a ``slice'' at the constant Galactic
longitude of \pulsar\ (351\fdg688) is shown in Figure~\ref{fig:cont}.
The total continuum emission toward PSR~B1718-35 in the GBT beam
is $9.04\pm0.05$\,K.  A two component
Gaussian fit that can be assumed to approximate the smooth Galactic synchrotron
and free-free emission contribution to the continuum emission was made.  
As can be seen in Figure ~\ref{fig:cont}, this approximates the smooth
background continuum emission from Galactic synchrotron
and free-free emission reasonably well.
This fit then sets the smooth Galactic component emission at 5.8\,K and the 
continuum in front of \pulsar\ as having 3.51\,K.  The continuum 
emission in front of the pulsar comes from three different source that
are slightly blended (this is easily seen in Figure~\ref{fig:spitzer}).
At a Galactic latitude lower than \pulsar\ is the \hii\ region 
G351.662+0.518 \citep{jayhii} and at a higher Galactic latitude is
the SNR G351.7+0.8 \citep{green, whiteoak}.  Fitting an additional
three Gaussians to this emission results in a 
brightness temperature of $4.1\pm0.2$\,K for G351.662+0.518, 
$4.2\pm0.2$\,K for SNR G351.7+0.8 and $2.9\pm0.2$K for the continuum
source in front of \pulsar.
Data from the Parkes 6\,cm survey of the southern Galactic plane
at 5009\,MHz \citep{haynes} shows that the continuum source in front
of \pulsar\ has 
a flux of $2.3\pm0.2$\,K for the \pulsar\ line of sight.  This then
gives a spectral index of $-0.21\pm0.2$ (using $S=\nu^\alpha$) 
which is consistent with the expected \hii\ region thermal emission 
spectral index of $-0.07$.

\subsection{Comparing The Pulsar ``On'' and ``Off'' OH Absorption Spectra}

For the two previous detections of OH absorption against pulsars, the
opacity of the absorption has been greater and the linewidths have 
been narrower against the pulsar than against the continuum background
\citep[see][]{snez, joel}.  Two viable explanations were put
forward in \citet{snez}: a) the absorption comes from OH clouds whose
angular size is larger than the observed scattering
size of the pulsar but smaller
than the angular size of the telescope beam;  or b) there are
extra OH clouds that are seen in absorption against the continuum 
background with the larger
telescope beam that are not seen along the pulsar line of sight.
In the former case, if the OH cloud seen in absorption against the 
pulsar could be spectroscopically isolated in the pulsar 'off' spectrum,
then the linewidths in the two spectra should be the same.  The differences
in the optical depth of the line could then be used to constrain the size
of the cloud.  In the later case, the OH cloud is expected to be built
from smaller ``cloudlets'' \citep{snez} and we would expect the 
linewidths and the opacities to be different between the pulsar absorption
and the continuum background absorption spectra.  Of course, both of these
scenarios could be operating at the same time.

In Figure~\ref{fig:narrow} we see that the 1665 and 1667\,MHz absorption
lines at $\sim 1.7$\kms\ have a narrow component and a broader
component.  Gaussian fits for a narrow and broad linewidth components to the
1667\,MHz background continuum absorption spectrum are presented in
Table~\ref{table:datawidenarrow}.  The broad component Gaussian fit
is shown in the top panel of Figure~\ref{fig:abscompare}.  Comparing the
results in Table~\ref{table:datawidenarrow} and Table~\ref{table:datapulsar},
we see that the narrow component has the same velocity and that the 
linewidths agree at the $1 \sigma$ level for the pulsar and continuum
background cases.  In the bottom panel of
Figure~\ref{fig:abscompare}, I show the opacities of the two narrow components
and the difference between them.  The opacity of the pulsar ``off'' 
spectra was determined using an effective continuum strength of 
2.1\,K (see below) which makes
the opacities equal.  As can be seen in Figure~\ref{fig:abscompare},
there is no discernible difference between the narrow component in the
pulsar ``off'' spectrum and the absorption against the pulsar.  Thus,
only the size of the OH cloud will play a role in any difference between
the opacity observed against the background continuum and against the
pulsar.

Since we know the opacity of the OH from the absorption against the
pencil-thin beam of the pulsar and there is information on the continuum
brightness, limits can be determined for the size of the OH absorption
cloud.  This is done using
\begin{equation}
e^{-\tau} = { I(\nu) \over f_\Omega T_{bg}}+1 = {I(\nu) \over T_{eff}}+1
\end{equation}
where $I(\nu)$ is the measured, baseline subtracted spectrum
(top panel of Figure~\ref{fig:abscompare}) and 
$f_\Omega = \Omega_{OH} / \Omega_{GBT}$ is
the size of the OH cloud ($\Omega_{OH}$) relative to the size of the telescope
beam ($\Omega_{GBT}$). 
$T_{eff} = f_\Omega T_{bg}$ is the effective brightness temperature
and corresponds to the OH cloud size weighted background brightness
temperature averaged over the entire GBT beam.  Performing a 
least-squares, non-linear fit to the observed narrow linewidth component
results in $T_{eff} = 2.1\pm0.2$\,K.  Allowing all of the continuum flux
to reside behind the OH cloud, $T_{bg} = 8.7\pm0.2$\,K where 5.8\,K is from
the smooth Galactic continuum and 2.9\,K is from the \hii\ region, gives
a minimum size of the cloud assuming that the cloud is spherically 
symmetric.  For this case $f_\Omega = 0.24\pm0.02$.
Using the 7.4 arc-minute FWHM size of the GBT beam, the minimum
OH cloud radius is $0.89\pm0.08$ arc-minutes which at a distance of
1.7\,kpc corresponds to $0.44\pm0.04$\,pc. 
If the OH cloud is behind the \hii\ region, and assuming that there
is little contribution to the Galactic continuum between the Sun and
\ngc, $f_\Omega = 0.36\pm0.04$ and the size of the OH cloud would is 
$1.3\pm0.1$ arc-minutes, which
corresponds to $0.64\pm0.06$\,pc at a distance of 1.7\,kpc. 
These size scales along with the column densities derived in 
\S~\ref{sec:column}  translate into densities of order $n_H \sim 10$ --
100\,cm$^{-3}$ for the OH cloud.

An ellipse with semi-major and semi-minor axes of 3 and 1.8 arc-minutes
roughly covers the area of 8~micron emission seen in Figure~\ref{fig:spitzer}.
If we assume that the OH cloud size is represented by the 8~micron
dust cloud size, then $f_\Omega \sim 0.36$.  This is the same number
derived above assuming that the OH cloud is behind the \hii\
region and in front of most of the Galactic background continuum emission.
It seems likely that the OH cloud lies behind the \hii\ region.

From the FWHM linewidths in Table~\ref{table:datawidenarrow}, 
the narrow linewidth component corresponds to a thermal temperature
$T < 830\pm240$\,K.  The broad linewidth components corresponds to
a thermal temperature of $T < 10400 \pm 1200$\,K.  Since the OH might
be expected to be in regions with temperatures in the range 10 -- 100\,K,
there is obviously a very large turbulent component to the linewidths.
The two components fit in very nicely with the molecular cloud (MC) --
\hii\ region interaction scenario.  The narrow linewidth OH could
be in the part of the MC that has not yet interacted with the \hii\
region while the broad linewidth component is in the part of the MC
interacting with the \hii\ region.

\subsection{Energy Input Into The Molecular Cloud}

If it is assumed that there is an interaction between the MC and an
\hii\ region, then we can estimate the amount of energy
input into the MC.  This is done by comparing the linewidths
of the narrow and broad components.  
From the above size (6 by 2.6 arcminutes or 2.97 by 1.28\,pc at
a distance of 1.7\,kc) and column density estimates 
($N_{OH} < 6.7 \times 10^{13}$\,cm$^{-2}$), there are
$<2.4\times 10^{51}$ OH particles in the narrow line component of the cloud.  
An OH linewidth of $1.5$\kms\ corresponds to an energy of 
$3.2\times 10^{-13}$\,ergs for each OH molecule, using $E={1/2} m v^2$.  
This gives a 
total energy of $<7.7\times 10^{38}$\,ergs for all the OH molecules
in the narrow linewidth component.  An OH linewidth of $5.3$\kms\ 
corresponds to an energy of 
$4.0 \times 10^{-12}$\,ergs for each OH molecule in the broad linewidth
component of the OH cloud.  The broad line-component has the same
characteristics between the 1720 and 1612\,MHz lines as does the narrow
linewidth component.  We can thus limit the column density of the broad
line to be $N_{OH} < 1.9 \times 10^{13}$\,cm$^{-2}$ which results
in a total energy of $<2.1\times 10^{39}$\,ergs for all the OH molecules
in the broad linewidth component.
This suggests that a few times $10^{39}$\,ergs of energy have been
input into the broad linewidth component.  

If we assume that the same amount of energy per particle has been
added to all molecular species in the broad component of the OH cloud,
then a total of a few times $10^{46}$\,ergs of energy have been 
deposited into the cloud.  SNR typically involve a release of 
$\sim 10^{51}$\,ergs of energy.  A cluster of O and B stars that
form \hii\ regions can also output
$\sim 10^{51}$\,ergs of energy over their lifetimes.
So it is not unreasonable to assume that the broad linewidth component
of the OH cloud has interacted with an \hii\ region or SNR.

\section{Conclusions}

\pulsar\ is only the third pulsar found to have OH absorption.  In all 
three cases the OH absorption arises in the
interaction of a MC with a SNR or an \hii\ region. Forty-seven (47)
pulsars have been searched for OH absorption.  This suggests that
OH absorption against pulsars is either quite rare, or that the 
absorption in most MCs is very weak and below the detection capabilities of
current telescopes without very deep searches.  That all known cases
involve a MC that is in the process of being destroyed by a SNR
or \hii\ region and that the detection rate is only $\sim 6$\%
implies that OH absorption against pulsars is a rare occurrence.

The detection rate using pulsar brightness is about $\sim 6$\%.  I found
a detection rate of 17\% using scintillation strength for the selection
criteria.  I also had a detection rate of 0\% using detected OH emission
that could be between the Sun and the pulsar.  Since the \pulsar\ OH detection
was found in a search of six pulsars based on their strong scattering,
and \pulsar\ is the 7th most strongly scattered pulsar,
some credence can be given to the hypothesis of \citet{boldyrev} but
it by no means provides conclusive proof for their idea that scattering
originates in PDRs at the boundaries of \hii\ regions and MCs. 

\begin{figure}
\epsscale{0.9}
\includegraphics[width=6.5in]{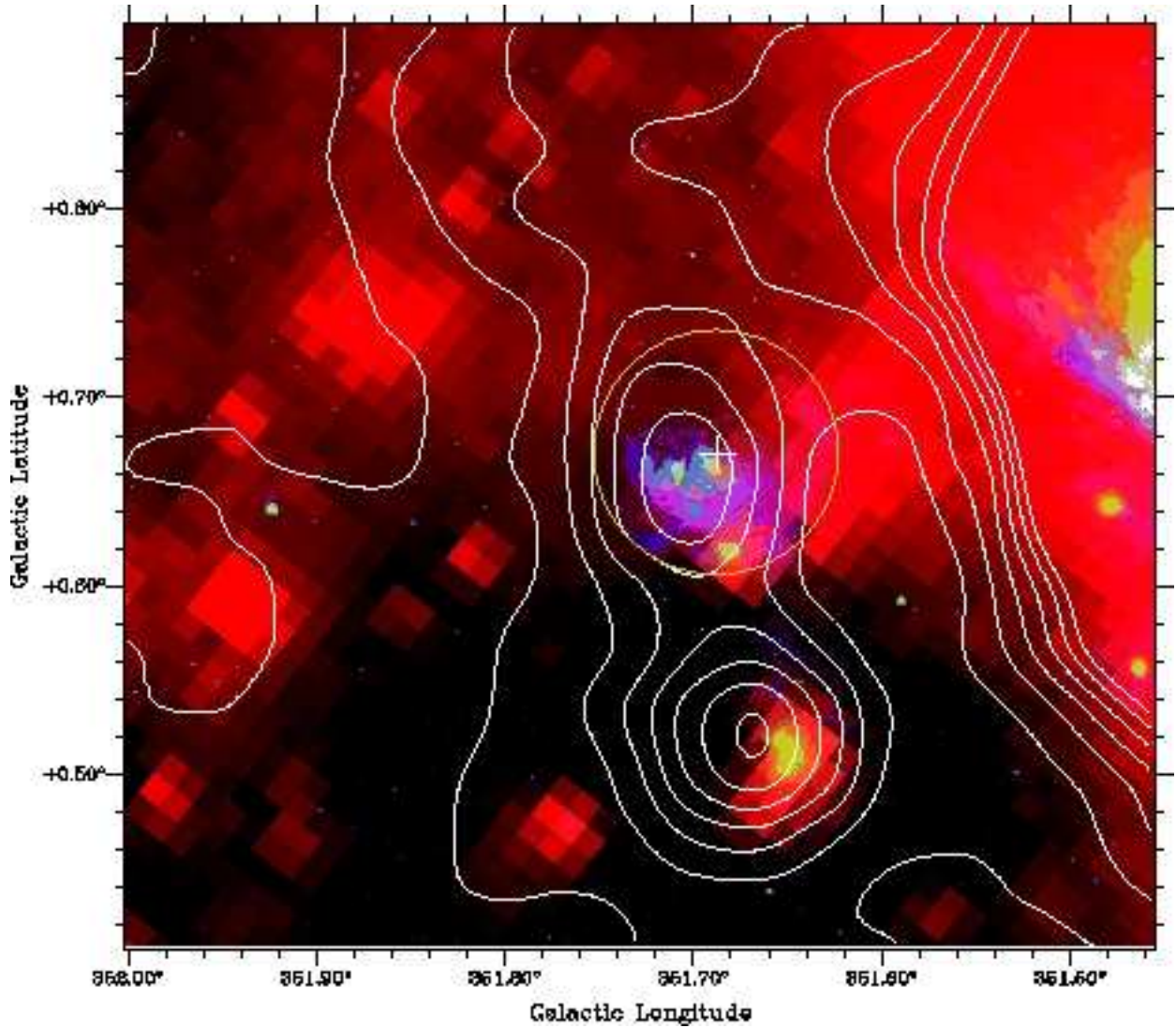}
\caption{The region around PSR~B1718-35.  This false color image is
comprised of Spitzer GLIMPSE~II 8 micron data (blue), Spitzer MIPS 24
micron data (green), SHASSA ${\rm H_\alpha}$ data (red) \citep{shassa}, 
and Parkes
6\,cm continuum emission data (white contours) \citep{haynes}.  
The contour levels
are from 0 to 5\,K every 0.5\,K.  The position of
PSR~B1718-35 is shown by the white cross.  The size of the GBT beam
at 1665\,MHz is shown by the green circle.  NGC~6334 is visible at
the right edge of the image.}
\label{fig:spitzer}
\end{figure}

\begin{figure}
\epsscale{0.9}
\includegraphics[scale=0.65, angle=90]{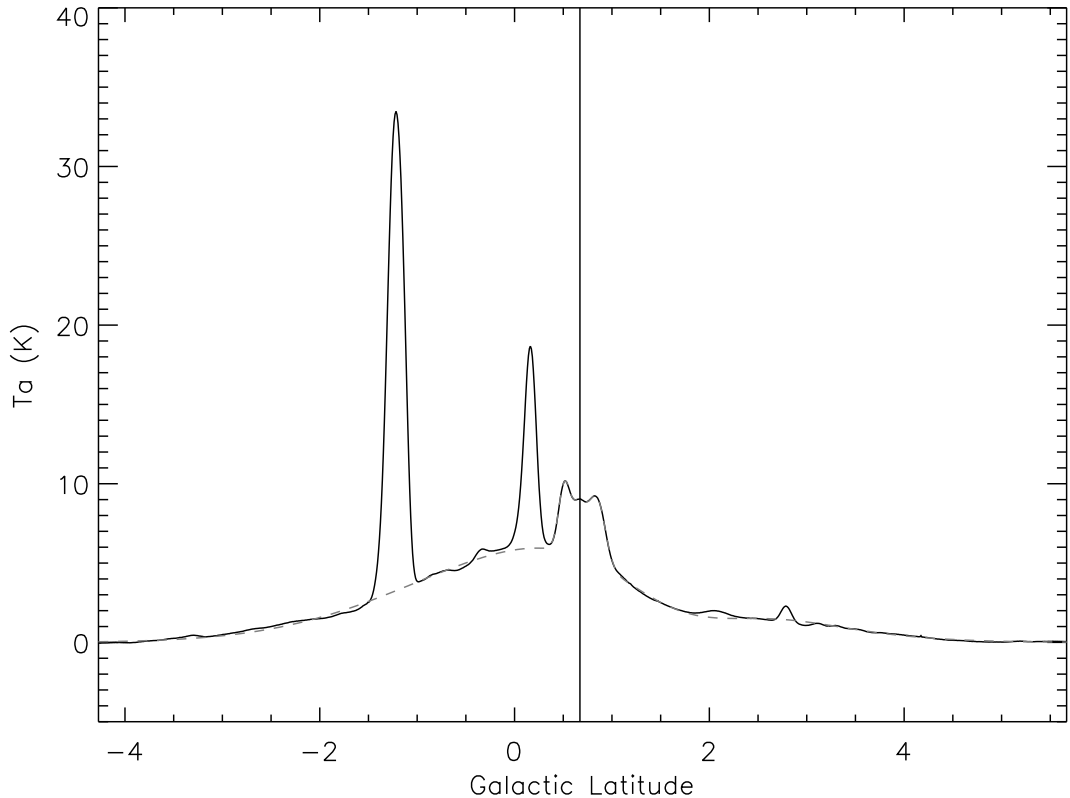}
\caption{Continuum data slice along Galactic longitude 351\fdg688 passing
through the location of PSR~B1718-35.  These data have been corrected for
atmospheric opacity ($\tau=0.0103$, $T_{atm}=250$\,K) and have had a 
constant zenith system temperature of 15\,K removed.  The vertical line
marks the position of PSR~B1718-35.  The dashed gray line is a two component
Gaussian fit that can be assumed to approximate the smooth Galactic synchrotron
and free-free emission contribution to the continuum emission along with three
more Gaussians fitted for the continuum sources near the PSR~B1718-35 line of
sight.  The total
continuum emission toward PSR~B1718-35 is 9.0\,K.  The smooth 
Galactic component is 5.8\,K and the continuum source along the PSR~B1718-35
line of sight is 2.9\,K.}
\label{fig:cont}
\end{figure}

\begin{figure}
\epsscale{0.9}
\includegraphics[scale=0.65, angle=90]{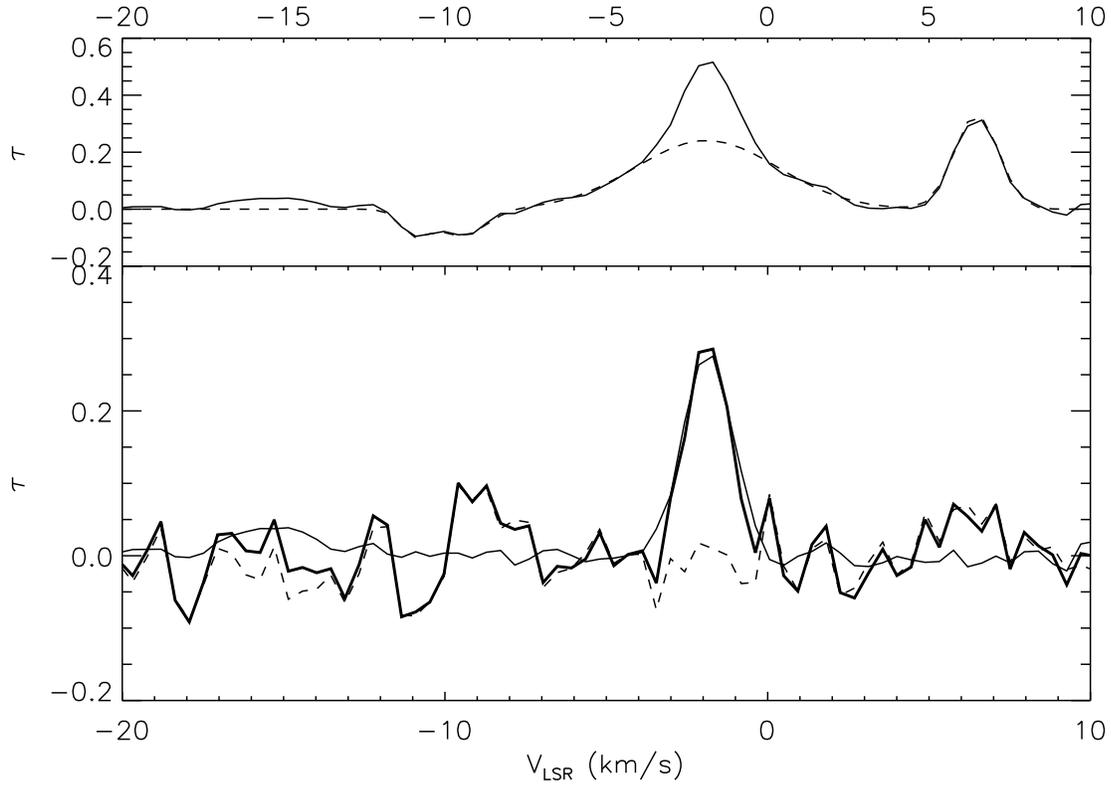}
\caption{Top panel: The opacity of the pulsar 'off' 1665\,MHz spectrum assuming
a continuum flux of 2.1\,K (solid line) and fitted Gaussians to the
broad component of the absorption line seen against the pulsar and 
other lines (dashed line).  Bottom Panel: Pulsar 1665\,MHz opacity
(thick solid line), the narrow component of the pulsar 'off' spectra after
subtracting the fit to the broad component (thin solid line) and the
differences between the spectra (dashed line).}
\label{fig:abscompare}
\end{figure}

In all likelihood, the continuum object along the line of sight
toward \pulsar\ is an \hii\ region associated with the
\ngc\ complex.  The spectral index of the source is consistent 
with what is expected for an \hii\ region.  
The morphology (see Figure~\ref{fig:spitzer}) 
of cold dust surrounding warm dust is not consistent with the
source being a SNR.  There is some H$_\alpha$ emission associated
with the continuum source (see Figure~\ref{fig:spitzer}).  
Unfortunately, the SHASSA survey \citep{shassa} does not provide
velocity information for the H$_\alpha$.  Although this line of
sight was observed as part of the WHAM survey \citep{wham}, the
1$^\circ$ resolution convolves the emission from the \hii\ region
toward \pulsar\ with other \hii\ regions in \ngc.
Thus it is not known if
the H$_\alpha$ arises at the same velocities as the OH absorption
against \pulsar.  Hydrogen
radio recombination lines or higher spatial resolution H$_\alpha$ with
velocity information observations should be performed to confirm that
the continuum source is an \hii\ region.

It was found that the OH cloud along the line of sight of \pulsar\
is likely behind the \hii\ region.  The OH cloud
has a broad linewidth component and a narrow linewidth
component.  Only the narrow linewidth component is seen in 
absorption against \pulsar.  A similar situation of broad and
narrow linewidth components with only the narrow component seen in absorption
against the pulsar was observed for \arecibo\ \citep{snez}.
This is also the case for \parkes\ \citep{joel}.
The opacities of the OH absorption against
\pulsar\ and against the continuum background (pulsar ``off'' spectrum)
can be reconciled via consideration of the beam-filling factor of the 
OH cloud.  This should be investigated further for the \arecibo\ and \parkes\
OH absorption.

\acknowledgements
This work is based [in part] on observations made with the Spitzer Space 
Telescope, which is operated by the Jet Propulsion Laboratory, California 
Institute of Technology under a contract with NASA.
The Southern H-Alpha Sky Survey Atlas (SHASSA) is supported by the National 
Science Foundation.

\clearpage
\begin{deluxetable}{lccll}
\tabletypesize{\scriptsize}
\tablecaption{Dates of the observations for the pulsars without 
detectable OH absorption with the GBT.\label{table:obs}}
\tablewidth{0pt}
\tablehead{
\colhead{Pulsar} & \multicolumn{2}{c}{Galactic Coordinates} & 
\colhead{Dates of Observations} & \colhead{Integration Time} \\
 & $l$ & $b$ & & (hrs) 
}
\startdata
\srca\ & $357\fdg099$ & $-0\fdg219$ & 2005 Jul 31 -- Aug 08 & 5.5 \\
\srcb\ &   $4\fdg257$ & $+0\fdg503$ & 2005 Aug 25 -- Aug 26 & 4.75 \\
\srcc\ & $12\fdg909$ & $ +0\fdg388$ & 2005 Aug 06 -- Aug 25 & 10.5 \\
\srcd\ & $16\fdg406$ & $ +0\fdg610$ & 2005 Jun 09 -- Aug 06 & 9.25 \\
\srce\ & $17\fdg160$ & $ +0\fdg482$ & 2005 Jun 01 -- Sep 01 & 18.25 \\
\srcf\ & $160\fdg363$ & $ +3\fdg077$ & 2006 May 14 -- Sep 30 & 36.75 \\
\srcg\ & $345\fdg718$ & $-0\fdg197$ & 2006 Jun 02 -- Aug 29 & 6.0 \\
\srch\ & $358\fdg295$ & $ +0\fdg238$ & 2006 Jul 21 -- Jul 30 & 3.75 \\
\srci\ & $25\fdg456$ & $ +4\fdg732$ & 2006 May 27 -- Oct 31 & 18.75 \\
\srcj\ & $23\fdg272$ & $ +0\fdg298$ & 2006 Jul 21 -- Jul 27 & 10.25 \\
\srck\ & $22\fdg263$ & $-1\fdg415$ & 2006 Jun 07 -- Jun 29 & 5.75 \\
\srcl\ & $31\fdg339$ & $ +0\fdg039$ & 2006 May 08 & 6.0 \\
\srcm\ & $86\fdg909$ & $-2\fdg012$ & 2006 Jun 19 -- Jul 25 & 8.25 \\
\srcn\ & $89\fdg003$ & $-1\fdg266$ & 2006 May 15 -- May 19 & 5.25 \\
\srco\ & $342\fdg457$ & $ +0\fdg922$ & 2006 May 29 -- May 31 & 4.5 \\
\enddata
\end{deluxetable}

\clearpage
\begin{deluxetable}{lc}
\tabletypesize{\scriptsize}
\tablecaption{Dates of observations detecting OH absorption toward
PSR~B1718-35 with the GBT.\label{table:obs1718}}
\tablewidth{0pt}
\tablehead{
 & \colhead{Bandwidth Per Band} \\
\colhead{Date}  & \colhead{(MHz)}
}
\startdata
2005 Aug. 24--29 & 2.5  \\ 
2005 Aug. 29--31 &  0.625  \\ 
2005 Oct. 08 & 0.625  \\ 
\enddata
\end{deluxetable}

\clearpage
\begin{deluxetable}{llrrll}
\tabletypesize{\scriptsize}
\tablecaption{Results of previous searches for OH absorption against pulsars.
The first column gives the pulsar observed.  The second column is the pulse 
broadening time found in the
ATNF pulsar catalog \citep{atnf} while
the third column gives the ranking in terms of the strength of the
pulse broadening for all pulsars.  The fourth column gives the integrated
CO intensity for all CO along the line of sight (not necessarily just
the CO that is foreground to the pulsar) using the survey data from \citet{co}.
The fifth column gives the
observed opacity or opacity limits where mentioned.
The sixth column 
gives the reference for the OH absorption observations.\label{table:scattering}}
\tablewidth{0pt}
\tablehead{
\colhead{Pulsar} & \colhead{${\rm \tau_{sc}^{1\,GHz}~(sec)}$} & \colhead{${\rm \tau_{sc}}$ Rank} & 
\colhead{${\rm W_{CO}~(K~km~sec^{-1})}$} & \colhead{$Opacity $} & \colhead{Reference}
}
\startdata
\arecibo\  & $1.51 \times 10^{-1}$ &   4 & 121.58 & 0.9 & \citet{snez} \\
\searcha\  & $9.33 \times 10^{-8}$ & 106 &     -- & --  & \citet{snez} \\
\searchb\  & $1.51 \times 10^{-5}$ &  71 &     -- & --  & \citet{snez} \\
\searchc\  & $3.89 \times 10^{-8}$ & 115 &   0.37 & --  & \citet{snez} \\
\searchd\  & $4.27 \times 10^{-8}$ & 114 &   0.05 & --  & \citet{snez} \\
\searche\  & $3.47 \times 10^{-5}$ &  65 &  90.86 & possible &  \citet{snez} \\
\searchf\  & $1.78 \times 10^{-9}$ & 142 &   0.03 & --  & \citet{snez} \\
\parkes\  & $1.12 \times 10^{-2}$ &  29 & 133.25 & 0.1 & \citet{joel} \\
\searchg\  & $7.76 \times 10^{-6}$ &  76 &   6.06 & $< 0.1$  & \citet{joel}\\ 
\searchh\  & $6.03 \times 10^{-5}$ &  62 &   0.00 & $< 0.1$  & \citet{joel}\\ 
\searchi\  & $7.59 \times 10^{-6}$ &  77 &   2.11 & $< 0.07$  & \citet{joel}\\ 
\searchj\  & --                    & --  &    --  & $< 0.1$  & \citet{joel}\\ 
\searchk\  & --                    & --  &    --  & $< 0.1$ & \citet{joel}\\ 
\searchl\  & --                    & --  &    --  & $< 0.1$ & \citet{joel}\\ 
\searchm\  & --                    & --  &    --  & $< 0.3$  & \citet{joel}\\ 
\searchn\  & --                    & --  &    --  & $< 0.1$ & \citet{joel}\\ 
\searcho\  & $2.88 \times 10^{-3}$ &  44 &   1.02 & $< 0.1$ & \citet{joel}\\ 
\searchp\  & $7.76 \times 10^{-6}$ &  76 &   6.06 & $< 0.1$ & \citet{joel}\\ 
\searchq\  & $4.47 \times 10^{-3}$ &  39 &  16.81 & $< 0.1$ & \citet{joel}\\ 
\searchr\  & --                    & --  &    --  & $< 0.2$  & \citet{joel}\\ 
\searchs\  & --                    & --  &    --  & $< 0.2$  & \citet{joel}\\ 
\searcht\  & $5.50 \times 10^{-7}$ &  88 &  55.67 & $< 0.05$  & \citet{joel}\\ 
\searchu\  & --                    & --  &    --  & $< 0.3$  & \citet{joel}\\ 
\searchv\  & $2.19 \times 10^{-7}$ &  96 &   9.89 & $< 0.3$ & \citet{joel}\\ 
\searchw\  & --                    & --  &    --  & $< 0.3$  & \citet{joel}\\ 
\searchx\ & $6.31 \times 10^{-8}$ & 111 & 0.08 & $< 0.1$ & \citet{galt} \\
\searchy\ & $2.57 \times 10^{-8}$ & 121 & 0.34 & $< 0.04$ & \citet{slysh} \\
\searchz\ & $2.88 \times 10^{-7}$ &  92 & -- & $< 0.15$ & \citet{slysh} \\
\searchaa\ & $7.76 \times 10^{-6}$ &  76 & 6.06 & -- & \citet{slysh} \\
\searchbb\ & $6.76 \times 10^{-8}$ & 109 & -- & -- & \citet{slysh} \\
\enddata
\end{deluxetable}

\clearpage
\begin{deluxetable}{lcccc}
\tabletypesize{\scriptsize}
\tablecaption{$ 1 \sigma$ Opacity upper limits for OH absorption toward the 
pulsar observations presented in this paper. Note 
that the GBT L-band receiver has a resonance around 1720~MHz which causes the
sensitivity at this frequency to be compromised, as one polarization does not 
contribute very much to the signal-to-noise.\label{table:limits}}
\tablewidth{0pt}
\tablehead{
\colhead{Pulsar} & \multicolumn{4}{c}{Opacity Upper Limits} \\
\colhead{} & \colhead{1665\,MHz} & \colhead{1667\,MHz} & \colhead{1612\,MHz} & \colhead{1720\,MHz}
}
\startdata
\srca\ & 0.07 & 0.06 & 0.08 & 0.1 \\
\srcb\ & 0.2 & 0.2 & 0.2 & 0.4 \\
\srcc\ & 0.06 & 0.06 & 0.06 & 0.10 \\
\srcd\ & 0.03 & 0.03 & 0.03 & 0.04 \\
\srce\ & 0.04 & 0.05 & 0.04 & 0.09 \\
\srcf\ & 0.06 & 0.06 & 0.06 & 0.07 \\
\srcg\ & 0.07 & 0.08 & 0.07 & 0.1 \\
\srch\ & 0.1 & 0.1 & 0.1 & 0.1 \\
\srci\ & 0.05 & 0.05 & 0.04 & 0.04 \\
\srcj\ & 0.2 & 0.2 & 0.2 & 0.7 \\
\srck\ & 0.08 & 0.08 & 0.07 & 0.09 \\
\srcl\ & 0.04 & 0.04 & 0.03 & 0.07 \\
\srcm\ & 0.09 & 0.1 & 0.1 & 0.1 \\
\srcn\ & 0.04 & 0.04 & 0.04 & 0.05 \\
\srco\ & 0.06 & 0.06 & 0.05 & 0.07 \\
\enddata
\end{deluxetable}

\clearpage
\begin{deluxetable}{cccccc}
\tabletypesize{\scriptsize}
\tablecaption{Gaussian fit results for the observed OH absorption against 
PSR B1718-35.\label{table:datapulsar}}
\tablewidth{0pt}
\tablehead{
\colhead{$\nu$} & \colhead{$\Delta \nu$} & \colhead{$\sigma_\tau$} & \colhead{$\tau$}         & \colhead{$V_{lsr}$} & \colhead{FWHM} \\   
 \colhead{(MHz)}     & \colhead{(MHz)}     & \colhead{} & \colhead{} & \colhead{(\kms )}     & \colhead{(\kms )}
}
\startdata
 1665 & 2.5 & 0.008 & $0.076 \pm 0.009$ & $-1.5 \pm 0.2$ & $3.0 \pm 0.4$ \\ 
 1667 & 2.5 & 0.009 & $0.074 \pm 0.008$ & $-2.0 \pm 0.2$ & $3.2 \pm 0.4$ \\ 
 1612 & 2.5 & 0.01 & $0.042 \pm 0.009$ & $-1.3 \pm 0.4$ & $ 3 \pm 1$ \\ 
 1720 & 2.5 & 0.01 & \nodata & \nodata & \nodata \\  
\\
1665 & 0.625 & 0.04 & $0.20 \pm 0.03$ & $-1.6 \pm 0.1$ & $1.6 \pm 0.3$ \\  
1667 & 0.625 & 0.04 & $0.30 \pm 0.03$ & $-1.86 \pm 0.07$ & $1.5 \pm 0.2$ \\ 
1612 & 0.625 & 0.04 & $0.14 \pm 0.03$ & $-1.7 \pm 0.1$ & $1.5 \pm 0.4$\\ 
1720 & 0.625 & 0.06 & \nodata & \nodata & \nodata \\  
\enddata
\end{deluxetable}

\clearpage
\begin{deluxetable}{ccc}
\tabletypesize{\scriptsize}
\tablecaption{Gaussian fit results for the observed 1667\,MHz OH absorption 
against the background continuum emission (pulsar ``off'' spectrum).  The
opacities were determined using an effective background continuum emission 
of 2.1\,K.
\label{table:datawidenarrow}}
\tablewidth{0pt}
\tablehead{
\colhead{$\tau$}         & \colhead{$V_{lsr}$} & \colhead{FWHM} \\   
\colhead{} & \colhead{(\kms )}     & \colhead{(\kms )}
}
\startdata
$0.30 \pm 0.02$ & $-1.86 \pm 0.03$ & $1.8 \pm 0.1$ \\  
$0.23 \pm 0.02$ & $-1.91 \pm 0.07$ & $5.3 \pm 0.3$ \\  
\enddata
\end{deluxetable}

\clearpage

\clearpage

\clearpage

\clearpage

\clearpage

\clearpage

\end{document}